\documentclass[aps,pra]{revtex4}
\usepackage{epsfig}
\newcommand{\be}{\begin{equation}}
\newcommand{\bea}{\begin{eqnarray}}
\newcommand{\ee}{\end{equation}}
\newcommand{\eea}{\end{eqnarray}}
\newcommand{\eq}[1]{Eq.~\ref{#1}}

\newcommand{\wzz}{W_{zz}^{[1/2]}}

\newcommand{\w}{\omega}
\begin{document}
\title{Decoherence induced by Smith-Purcell radiation}
\author{Ezequiel Alvarez and Francisco D.\ Mazzitelli}

\affiliation{ Departamento de F\'\i sica {\it J.J. Giambiagi}, FCEyN UBA,
Ciudad Universitaria, Pabell\' on I, 1428 Buenos Aires, Argentina}

\begin{abstract}

{The interaction between charged particles and the vacuum
fluctuations of the electromagnetic field induces decoherence, and
therefore affects the contrast of fringes in an interference
experiment. In this article we show that if a double slit
experiment is performed near a conducting grating, the fringe
visibility is reduced. We find that the reduction of contrast is
proportional to the number of grooves in the conducting surface,
and that for realistic values of the parameters it could be large
enough to be observed. The effect can be computed and understood
in terms of the Smith-Purcell radiation produced by the surface
currents induced in the conductor.}

\end{abstract}

\pacs{PACS number(s):03.65.Yz,03.75.Dg}

\maketitle

\section{Introduction}

The interaction of a quantum system with its environment produces
decoherence, which is one of the main ingredients to understand
the quantum to classical transition \cite{zurek}. If an
interference experiment is performed with charged particles, the
unavoidable interaction with the vacuum fluctuations of the
electromagnetic field induces decoherence, and therefore affects
the visibility of the fringes \cite{bp}-\cite{mpv}. Indeed, in
previous papers it has been shown that the presence of a perfectly
conducting plane surface has a small influence on the contrast of
the fringes. This was originally pointed out in Ref. \cite{ford},
and the results were subsequently revised and generalized in
Ref.\cite{mpv}. The interaction with the vacuum fluctuations of
the electromagnetic field can enhance or suppress the contrast
with respect to the visibility in vacuum (absence of conducting
plane), depending on the relative orientation of the conducting
plane and the plane of the trajectories of the particles
\cite{mpv}. The modification of the visibility of the fringes can
be understood in very simple terms \cite{imry}: if an electron
moves with a typical frequency $\Omega$ along the trajectory of
the experiment, and if the two trajectories are separated by a
distance $R$, the fringes visibility decays a factor $(1-P)^2$,
where $P$ is the probability that a dipole $p=eR$ oscillating at a
frequency $\Omega$ emits a photon. If the experiment is performed
in front of a conducting plane, the image charges enhance the
effect when the conductor is perpendicular to the plane of the
experiment. Alternatively, the image charges suppress the
decoherence  when the trajectories are contained in a plane
parallel to the conductor.   In principle, a similar effect is
present for neutral particles with electric and/or magnetic dipole
moment \cite{mpv}.

As $P$ is proportional to $e^2 v^2$, where $v$ is the mean
velocity of the particles with charge $e$, this effect is
extremely  small, and therefore very difficult to observe, unless
the charged particles are made to oscillate many times before
reaching the screen \cite{mpv}. There are variants of the above
mentioned configurations in which the decoherence is notably
larger. For example, it has been argued that if a double slit
experiment is performed over a non-perfect conductor, the image
charges in the material will dissipate energy through Joule
heating, and will also affect the visibility of the fringes
\cite{anglin}.   This effect has been further analyzed in
Refs.~\cite{mach,levinson}, where the interaction of the
interfering wave packets with the electrons in the macroscopic
body has been modelled using different approaches.  A recent
experiment has confirmed that the presence of a semiconducting
plate induces decoherence \cite{sonnen}, although  the agreement
between theory and experiment is still not perfect. Another
possibility to control the decoherence in a double slit-like
experiment is to perform it in the presence of a classical
electromagnetic  field \cite{ford2,lmv}. When this is sufficiently
intense, random phases coming from the emission time of the wave
packets can also induce a large amount of decoherence.

In this paper we will explore another configuration that will also
prove to be potentially interesting from an experimental point of
view. When a charged particle moves with constant velocity over a
perfectly conducting grating, the currents induced in the surface
of the conductor produce radiation, that can be intuitively
understood as coming from the oscillating image charge. This is
the so called Smith-Purcell effect \cite{spe}. We argue that,
because of this radiation, if a double slit-like experiment is
performed near a conducting grating, there will be a suppression
of the interference fringes. Moreover, we expect the effect to be
enhanced by the number of grooves in the conductor, and therefore
it is more likely to be observed than previous proposals in Ref.
\cite{ford,mpv}.

The paper is organized as follows. In the next Section we present
the main formalism and a formula which, in principle, gives the
decoherence factor for a double slit experiment performed near a
conducting surface of arbitrary shape. We follow closely
Ref.\cite{mpv}. In Section 3 we evaluate the decoherence factor
for the particular case of  a conductor with a periodic grating.
We argue that, under reasonable assumptions, the reduction in the
visibility of the fringes can be estimated  using a quasi
equivalent system formed by the original charges and their images
in vacuum. We show that the decoherence factor is proportional to
the number of grooves. In Section 4 we discuss the experimental
feasibility of our proposal and include our main conclusions. Some
details of the calculations are relegated to the Appendix.  Along this work we use natural units, $\hbar=c=1$.

\section{Fringe visibility in a double slit experiment with boundary conditions}

Let us consider a double slit experiment performed with charged
particles (electrons). Assume that the two electron wave packets
follow well defined trajectories $\vec{X}_1(t)$ and $\vec{X}_2(t)$
that coincide at initial ($t=0$) and final ($t=T$) times. In the
absence of environment, the interference depends on the relative
phase between both wave packets at $t=T$. Because of the
interaction with the quantum electromagnetic field, the
interference pattern is affected. We assume an initial state of
the combined particle--field system of the form $|\Psi(0)\rangle=
(|\phi_1\rangle+|\phi_2\rangle)\otimes|E_0\rangle$. Here
$|E_0\rangle$ is the initial (vacuum) state of the field and
$|\phi_{1,2}\rangle$ are two states of the electron that are
localized around the initial point and that in the absence of
other interaction continue to be localized along the trajectories
$X_{1,2}(t)$ respectively. At later times, due to the particle
field interaction the state of the combined system becomes
\begin{equation}
|\Psi(t)\rangle= (|\phi_1(t)\rangle\otimes |E_1(t)\rangle+|\phi_2(t)\rangle
\otimes|E_2(t)\rangle).\label{state}
\end{equation}
Thus, the two localized states $|\phi_1(t)\rangle$
and $|\phi_2(t)\rangle$ become correlated with two different states of the field. Therefore, the
probability of finding a particle at a given position turns out to be
\begin{eqnarray}
Prob(\vec X,t)&=&|\phi_1(\vec X,t)|^2+|\phi_2(\vec X,t)|^2 \nonumber\\
&+& 2{\rm Re}(\phi_1(\vec X,t)\phi_2^*(\vec X,t)\langle E_2(t)|E_1(t)\rangle).
\label{probability}
\end{eqnarray}

The absolute value of the overlap factor $F=\langle E_2(t)|E_1(t)\rangle$ is responsible
for the decay in the fringe contrast,
which is the phenomenon we will analyze here. Calculating it
is conceptually simple since $F$ is nothing but the overlap between
two states of the field that arise from the vacuum under
the influence of two different sources (this factor is identical to the Feynman--Vernon
influence functional \cite{Feynman-Vernon}). Each of the two states of the field
can be written as
\begin{equation}
|E_m(t)\rangle=T\left(\exp(-i\int d^4x J_m^\mu (x)A_\mu (x))\right)|E_0\rangle,
\label{Ea}
\end{equation}
where $J_m^\mu (x)$ is the conserved $4$--current generated by the
particle following the classical trajectory $\vec X_m(t)$, i.e.
$J_m^\mu (\vec X,t)=\left(e,e\dot{\vec X}_m
(t)\right)\times\delta^3(\vec X-\vec X_m(t))$, ($m=1,2$). Using
this, the simplest way to derive an expression for the overlap is
based on the observation that as the QED action is quadratic in
the fields, the overlap must be a Gaussian functional of the two
currents $J_1$ and $J_2$. Thus, we can write the most general
Gaussian functional ansatz for $F$ as
\begin{eqnarray}
F&=&\exp(-i\int\int d^4x_1d^4x_2 J_m^\mu(x_1)G_{\mu\nu}^{mn}(x_1,x_2)
J_n^\nu(x_2))\nonumber\\
&\times&\exp(-i\int d^4x J_m^\mu(x) C_\mu^n(x)).
\label{FGaussian}
\end{eqnarray}
where a summation over the indices $m,n=1,2$ is implicit.
On the other hand, we can explicitly write down the expression for the overlap as
\begin{eqnarray}
F&=&\langle E_0| \tilde T\left(\exp(i\int d^4x J_2^\mu (x)A_\mu (x))\right)\nonumber\\
&\times&T\left(\exp(-i\int d^4x J_1^\mu (x)A_\mu (x))\right)|E_0\rangle.
\label{Foverlap}
\end{eqnarray}
The kernels $G_{\mu\nu}^{mn}$ and $C_\mu^m$ appearing in (\ref{FGaussian})
can be determined by identifying the functional derivatives of equations
(\ref{FGaussian}) and (\ref{Foverlap}). In this way, one can relate $C_m$ and $G_{mn}$
with the one and two point functions of the field operators. As we are
only interested in the absolute
value of the overlap, we will only present the result for this quantity here.
Denoting $|F|=\exp(-W)$, we get
\begin{equation}
W={1\over 2}\int d^4x\int d^4y (J_1-J_2)^\mu(x)D_{\mu\nu}(x,y)(J_1-J_2)^\nu(y),
\label{Wcharges}
\end{equation}
where $D_{\mu\nu}$ is the expectation value of the
anti--commutator of two field operators:
\begin{equation}
D_{\mu\nu}(x,y)={1\over2}\langle\{A_\mu(x),A_\nu(y)\}\rangle.
\nonumber
\end{equation}

 From the above derivation, it is clear that the the square of the overlap has
a simple physical interpretation:  $|F|^2$ is
equal to the probability for vacuum persistence in the presence of
a source $j_{\mu}=(J_1-J_2)_\mu$, which corresponds to
a time dependent electric dipole $\vec p={e}(\vec X_1(t)-\vec X_2(t))$ \cite{imry}.
This is of course not a physical dipole but a fictitious one,
formed by the difference of the currents  associated to the
wave packets that follow each interfering trajectory.
The information about the existence of a conducting shell is encoded in
the two-point function of the electromagnetic field.

The overlap factor has been computed in Ref.\cite{mpv}, both in
vacuum and in the presence of a conducting plane, by considering
the appropriate two-point functions. If we denote by $W_0$ the
result in empty space and by $W_{plane}$ the result in the
presence of a conductor,  the overlap factor is given by
$W_{plane}\approx 0$ when the trajectories are parallel to the
conductor (recoherence) and by $W_{plane}\approx 2 W_0$ when the
trajectories are perpendicular to the conducting surface
(decoherence). These results, valid when all trajectories are very
close to the conductor, can be reproduced using the method of
images and  taking into account that decoherence in empty space is
related to the probability of photon emission for the fictitious
varying dipole $\vec p$. When the conducting plane is parallel to
the dipole, the image dipole is $\vec p_{im}=-\vec p$. Therefore
the total dipole moment vanishes, and so does the probability to
emit a photon in the leading order approximation. The image dipole
cancels the effect of the real dipole and this produces the
recovery of the fringe contrast with respect to the empty space
case. On the other hand, when the conductor is perpendicular to
the trajectories, the image dipole is equal to the real dipole
$\vec p_{im}=+\vec p$. Therefore, the total dipole is twice the
original one. This in principle would lead us to conclude that the
total decoherence factor  should be four times larger than $W_0$.
However, in the presence of a perfect mirror photons can only be
emitted in the $z\geq 0$ region. This introduces an additional
factor of $1/2$ that gives rise to the final result. The
conclusion is that one can replace the original problem by an
equivalent one in which the conducting plane is replaced by the
image charges. The overlap factor can be computed using the
two-point function of the electromagnetic field in empty space, by
including the appropriate image charges. In the next Section, with
the aim of seeking an enhanced observable which could be measured
in present experiments, we will use this equivalence to compute
the overlap factor in a more complex geometry.

\section{Induced decoherence in trajectories above a grating conductor}

We will now analyze an interference experiment in which a charged
particle moves above a conducting medium with periodic grating
along two possible different paths. We expect that, within the
proper approximations, the radiation emitted by the accelerated
image charges leave imprinted the {\it which-path} information in
the electromagnetic field, inducing in this way decoherence and,
therefore, a loss of contrast in the interference fringes.

\begin{figure}[ht]
\begin{center}
$\begin{array}{c@{\hspace{1in}}c}
\multicolumn{1}{c}{\framebox{ \epsfxsize=.47\textwidth \epsffile{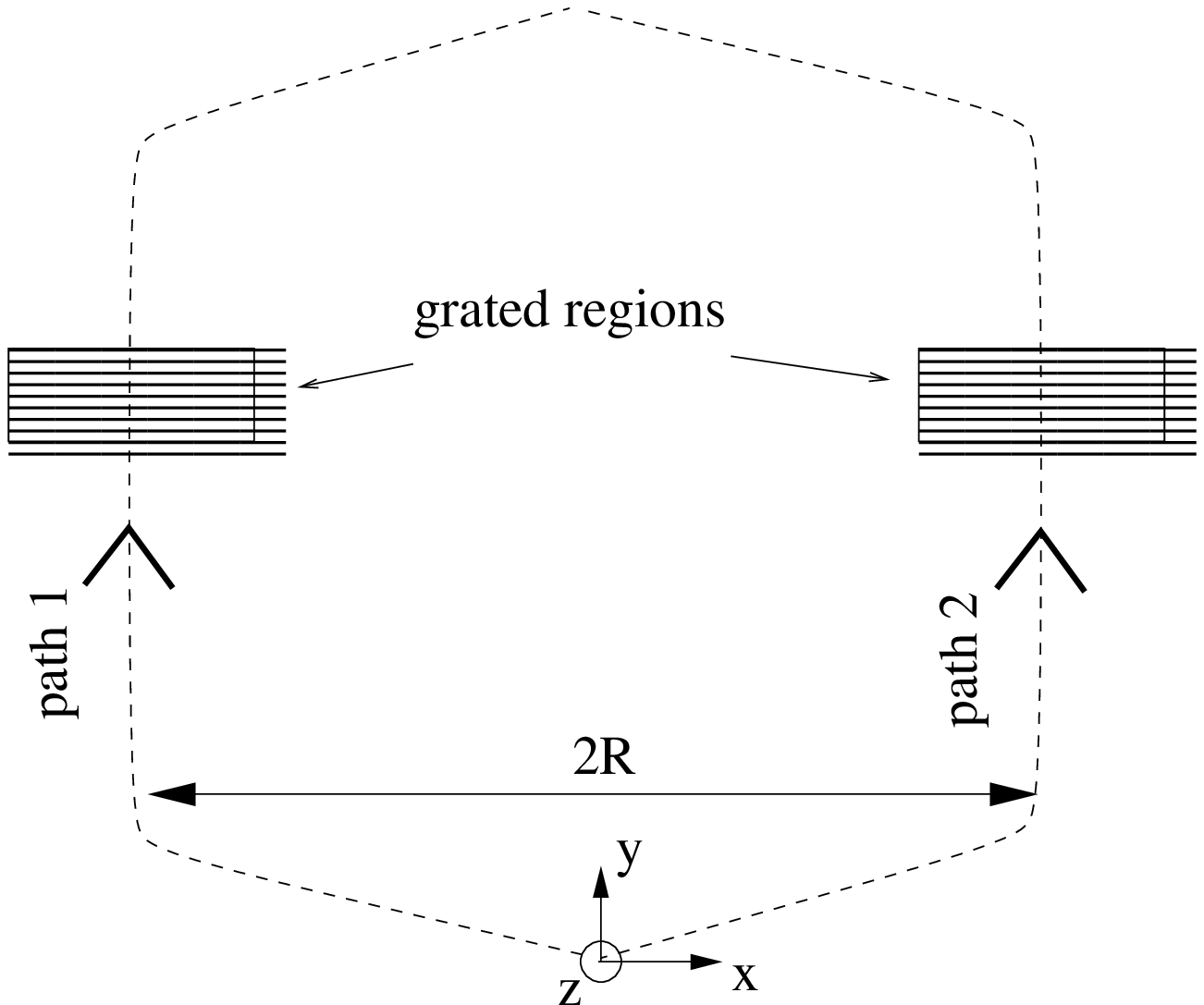} }}
&
\multicolumn{1}{c}{\framebox{ \epsfxsize=.473\textwidth \epsffile{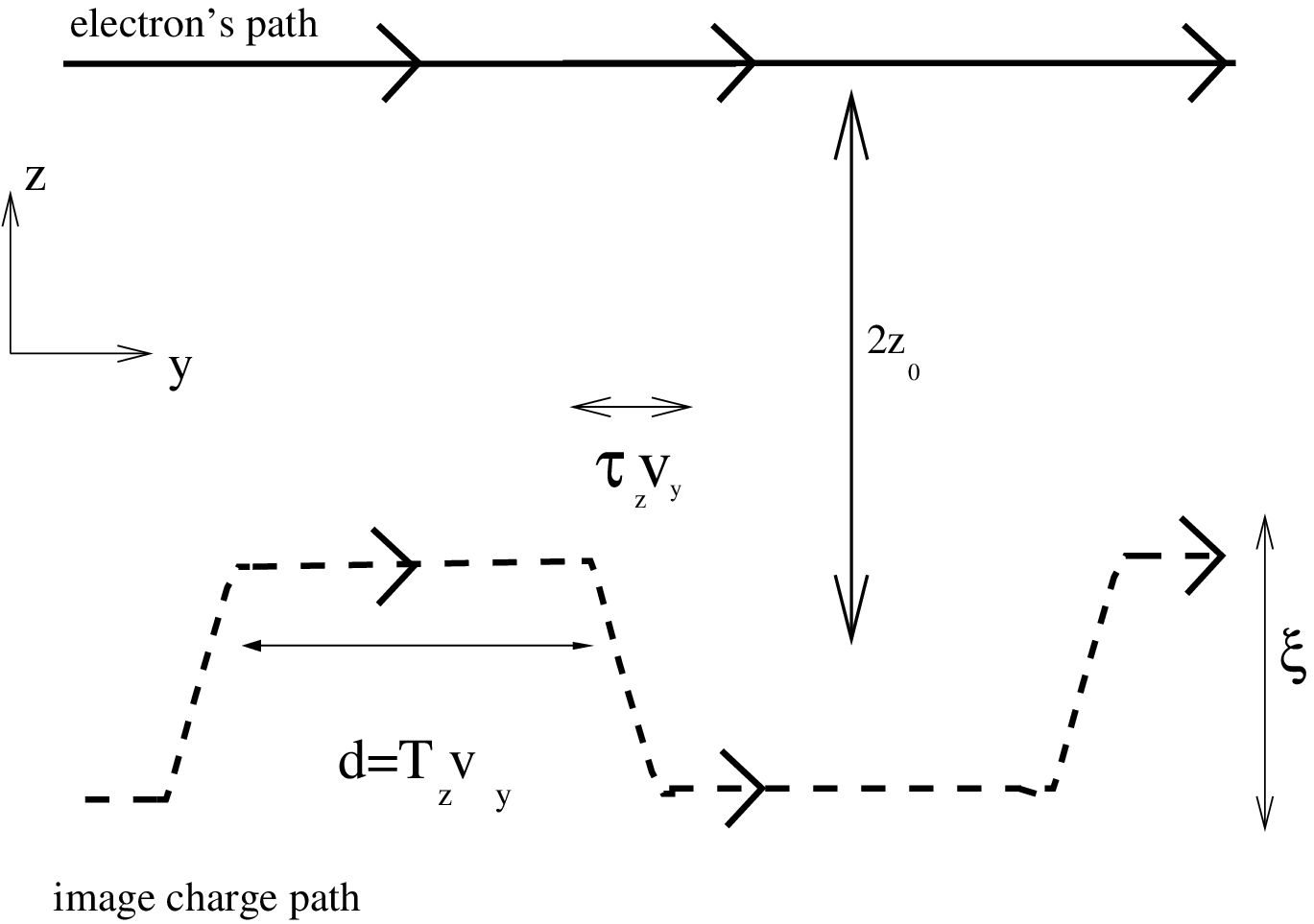} }}
 \\ [0.2cm]
\multicolumn{1}{c}{ \mbox{ \bf (a)}}  & \multicolumn{1}{c}{ \mbox{ \bf (b)} }
\end{array}$
\end{center}
\caption{$(a)$ General up-view of the two possible interfering
paths (not to scale).  The conductor is located behind the paths
from this view.  $(b)$ Side view of the charge path and the image
charge path in a region with grating in the conductor's surface.}
\label{setup}
\end{figure}

We consider an experimental setup in which an electron moves in
two possible trajectories above a conducting plate with a periodic
grating.  The grating shape below each trajectory has no need of
being equal to each other.  In particular, we study here the two
trajectories in Figure \ref{setup}a, where the electron travels at
a height $z_0$ above the conductor's surface, and the conductor's
surface consists in a grating of spatial period $2d$ and groove's
depth $\xi$. In order to calculate the overlap factor, one should
be able to compute the two-point function of the electromagnetic
field appearing in Eq. \ref{Wcharges}, in the presence of the
grating. Alternatively, one could model the currents induced in
the conducting surface and use the empty space two-point function.
Indeed, this approach has been considered in the literature in
order to calculate the radiated intensity distribution produced by
an electron beam passing over a metallic grating
\cite{Brownell98}.  A third possibility, when the geometry of the
conductor allows it, is to use the method of images to reproduce
the effects of the surface currents.

In this article, without loosing the sought effect and for the
sake of simplicity, we analyze the problem as if an electron's
image charge is placed below the conducting plate surface and
performs a trajectory with non-uniform vertical velocity $v_z$
which depends on the grating shape, as shown in Figure
\ref{setup}b. This could be achieved, for instance, if the grating
behaves as a periodic array of horizontal "infinite planes"
connected by sloping surfaces, giving the usual image charge
everywhere but on the position of the slopes, where we model the
surface current as an image charge with non vanishing vertical
velocity, going from one plane to the next one. To validate this
simplified model we will assume $d >> z_0, \xi$.  We may refer to
this as a {\it proximity approximation}. Moreover, to avoid the
problem of non physical velocities $v_z>1$ for the image charges,
we will assume that the slopes are not very steep. The  method of
images has already been considered to compute the intensity
distribution of the Smith-Purcell radiation \cite{Ishiguro}, as
well as to model the induced retarded currents in the conductor
\cite{Brownell98}.

We argue that, within this proximity approximation, the
decoherence factor corresponding to the experimental setup of the
charge moving in two possible paths above the conducting surface
equals one half of the decoherence factor produced by two charges
(one equal to the real charge and other equal to its image) in
vacuum which travel in two possible paths each one of them.   The
$1/2$ factor comes from the double counting of radiation in the
quasi-equivalent setup.  The two currents which interfere in this
scheme are
\begin{eqnarray}
J_1 &=& J_1^a + J_1^b  \label{J1}\\
J_2 &=& J_2^a + J_2^b   \label{J2}
\end{eqnarray}
with
\be
\left.
\begin{array}{rcl}
J_{1,2}^a (x) &=& e (1,\pm \dot s (t) \hat x + v_y \hat y) \, \delta^3 \left( \vec x - (\pm s(t)\hat x + v_y t \hat y + z_0 \hat z)\right) \\
J_1^b (x) &=& -e (1,\dot s (t) \hat x + v_y \hat y + \dot u (t) \hat z) \delta^3 \left( \vec x - (s(t)\hat x + v_y t \hat y +(- z_0 + u(t)) \hat z)\right)  \\
J_2^b (x) &=& -e (1,- \dot s (t) \hat x + v_y \hat y + \epsilon
\dot u (t) \hat z) \delta^3 \left( \vec x - (-s(t)\hat x + v_y
t\hat y +(- z_0 + \epsilon u(t)) \hat z)\right) .
\end{array}
\right. \label{currents} \ee The superscripts $a$ and $b$ stand
for {\it above} and {\it below}, respectively (the {\it above}
currents correspond to the physical charges, while  the {\it
below} currents describe the image charges),
 $s(t)$ is the
electron's trajectory in the $\hat x$-direction, $u(t)$ is the
trajectory in the $\hat z$-direction for the image charge, and
$\epsilon=0,\pm1$ depending on the relative shape of the gratings
in the conductor below each trajectory in the original setup.
$\epsilon=0$ means that below one of the trajectories there is a
grating and below the other the conductor is flat; $\epsilon=+1$
means that the perturbations in the conductor's surface (relative
to a flat surface) in each path are exactly the same; and
$\epsilon=-1$ means that the perturbations in each path are
opposite, i.e., if at a given point in one of the paths
 there is a {\it valley} in the perturbations, then at
the equivalent point in the other path there is a {\it peak}.
These particular values for $\epsilon$ will allow us to explore
the effect of different gratings below each path of the original
setup.

Defining \be W^{cd} = \frac{1}{2} \int d^4x d^4x' (J_1^c(x) -
J_2^c(x) )_\mu D^{\mu\nu}(x,x')  (J_1^d(x') - J_2^d(x') )_\nu ,
\ee with $c,d = a,b$, it is straightforward to obtain the
decoherence factor for the original setup that, in virtue of the
emitted Smith-Purcell radiation, will be denoted by $W_{SP}$: \be
W_{SP} = \frac{1}{2} (W^{aa} + W^{ab} + W^{ba} + W^{bb}),
\label{W1} \ee where $W^{aa}$ is the contribution that corresponds
to the interaction between the {\it above} currents, $W^{ab}$ to
the interaction of the {\it above} currents with the {\it below}
currents, and so on. The two-point function of the electromagnetic
field is the {\it free} two-point function, that in the Coulomb
gauge reads \be D_{ij}(x,x')=\int \frac{d^3k}{(2\pi)^3}\frac{1}{2
\omega }\left(\delta_{ij}-\frac{k_ik_j}{\omega^2}\right )e^{i\vec
k.(\vec x-\vec x')}\cos(\omega(t-t')) \ee where $i,j=x,y,z$.  By
computing explicitly the different terms in \eq{W1} it is easy to
obtain that, for instance, $W^{aa} = W_0$ (the result for empty
space produced by the currents above the conductor). Moreover, if
we define \be W^{cd}_{ij} = \frac{1}{2} \int d^4x d^4x' (J_1^c(x)
- J_2^c(x) )_i D^{ij}(x,x') (J_1^d(x') - J_2^d(x') )_j,
\label{Wcdij} \ee with no sum on $i$ and $j$, in such a way that
$W^{cd} = \sum_{i,j} W^{cd}_{ij}$, it is easy to see that \be
W_{plane} = \frac{1}{2} \sum_{c,d} \sum_{i,j\neq z} W^{cd}_{ij} .
\ee That is, if we do not include the $z$-component of the
currents in the computation of $W_{SP}$, we retrieve the
decoherence factor for a plane conductor. Therefore, the
decoherence factor due to the grating in the conductor equals the
factor associated to a plane conductor plus a modification, \be
W_{SP} = W_{plane} + \Delta W , \ee where $\Delta W$ comes from
the terms containing a non-vanishing $z$-component of the
velocity, \bea \Delta W &=& \frac{1}{2} W^{bb}_{zz} + \left(
W^{ab}_{xz} + W^{bb}_{xz} + W^{ab}_{yz} + W^{bb}_{yz} \right)  .
\label{dw1} \eea

Suppose now that the grating in the conductor has $N$ grooves,
such that the vertical velocity $\dot u(t)$ of the image charges
can be modelled as \bea \dot u(t) &=& \sum_{n=0}^{2N-1} (-1)^n \,
\dot u_0 (t - n T_z), \label{b} \eea where $\dot u_0(t)$
corresponds to the velocity shape each time there is a {\it steep}
in the conducting surface, $T_z$ is the time it takes the charge
to go from one steep to the next, $T_z = d / v_y$, and the
$(-1)^n$ factor accounts for the upwards and downwards steeps.
(Observe that $\dot u_0 (t)$ is different from zero only in
$t\in[0,\tau_z]$, where $\tau_z$ is the time it takes the charge
to go over the steep.)  Once we have modelled $\dot u(t)$ as the
sum in \eq{b}, we realize that the different terms in the RHS in
\eq{dw1} will contain either one of these sums ($W^{ab}_{xz},\
W^{bb}_{xz},\ W^{ab}_{yz}$ and $W^{bb}_{yz}$) or a modulus square
of this sum ($W^{bb}_{zz}$), depending on how many times the
$z$-component of the below currents is present in them. For
example
\bea W^{bb}_{zz} &=& e^2 \int \frac{d^3 k}{(2\pi)^3}
\frac{1}{\w} (1-\frac{k_z^2}{\w^2}) \left| \frac{e^{i k_x R} -
\epsilon \, e^{-i k_x R}}{2} \right|^2 \left| \int_0^{\tau_z} dt
\,
\dot u_0 (t) \, e^{i\w t} \right|^2 \cdot \nonumber \\
&& \left( \sum_{n,m=0}^{2N-1} (-1)^{n+m} \, e^{i(k_y v_y + \w) T_z
(n-m)}\right) . \label{hola0} \eea The modulus square in
$W^{bb}_{zz}$ gives rise to $N$ terms in which the phases in the
sum in \eq{hola0} are exactly cancelled and, hence, they are
relatively enhanced in comparison to the contribution coming from
the remaining $N^2-N$ terms.  Indeed, these other terms with
$n\neq m$ correspond to the interaction of the currents in the
$n$-th and $m$-th steeps, and we expect them to vanish for $T_z$
(or equivalently $d$) large enough.  In fact, this is the case,
since for $T_z >>\tau_z$ the phase factor's frequency as a
function of $\w$ is at least $T_z$, whereas the $\dot u_0 (t)$'s
Fourier transform has a typical width of $1/\tau_z$, therefore the
$\w$-integration gives a strong suppression if this Fourier
transform is well behaved.  If $\epsilon \neq 0$ then the
contribution of these $n\neq m$ terms also depends on the
relationship between $R$ and $T_z$, since for $2R > T_z$ we may
have that the current coming from the $n$-th steep in one path
interacts with the current coming from the $m$-th steep in the
other path.  The detailed analysis of this discussion is relegated
to the Appendix, where we show that under the reasonable
conditions $\tau_z \ll T_z$  (and $2R<T_z$ for $\epsilon \neq 0$),
the following approximation hold, \be W^{bb}_{zz} \approx 2N \wzz.
\label{uno} \ee Here $\wzz$ is the same as $W^{bb}_{zz}$ but
replacing $\dot u (t) \to \dot u_0(t)$ in the expression for the
$z$-current, \eq{currents}, i.e., it corresponds to the
decoherence factor of only half oscillation in the $z$-direction.
Under the same conditions, one can show that \be W^{bb}_{zz} \gg
W^{ab}_{xz} + W^{bb}_{xz} + W^{ab}_{yz} + W^{bb}_{yz} ,
\label{dos} \ee i.e. the leading contribution to the decoherence
comes from the term which is quadratic in the vertical component
of the velocity (see Appendix).

Integrating $\vec x$ and $\vec x'$ in \eq{Wcdij} it is
straightforward to obtain
\be \wzz = e^2 \int \frac{d^3
k}{(2\pi)^3} \frac{1}{\w} (1-\frac{k_z^2}{\w^2}) \left| \frac{e^{i
k_x R} - \epsilon \, e^{-i k_x R}}{2} \right|^2 \left|
\int_0^{\tau_z} dt \, \dot u_0 (t) \, e^{i\w t} \right|^2,
\label{W1osc} \ee
where we have used $v_y \ll 1$ and $k_z u(t)
\approx 0$ since, for simplicity, we are assuming a non-relativistic regime.

To summarize, we have found that for a double slit-like experiment
above a flat conducting plate, if the flat surface is changed to a
grating surface with $N$ grooves, then the decoherence factor is
enhanced as \be W_{plane} \longrightarrow W_{SP} \approx W_{plane}
+ 2N \wzz . \label{main} \ee This is the main result of this
article. Notice that this effect is due to the acceleration of the
image charge at the sloping surfaces, hence the effect is
proportional to $N$ and, as far as $z_0<<d$, is independent of
$d$.

An estimation of the enhancement in \eq{main} needs a calculation
of $\wzz$.  The precise value of $\wzz$, however, needs the
function $\dot u_0 (t)$, the distance between both paths, $R$, and
a value for $\epsilon$.  Nevertheless, we may study its generalities and then examine a particular example with a given function $\dot u_0 (t)$.

As a first observation notice that, in the case $\epsilon=0$, the
$\wzz$ decoherence factor is independent of $R$. This is to be
expected, since the Smith-Purcell radiation is in this case
entirely due to the image charge that corresponds to the
trajectory  that lies on the grating. For $\epsilon=\pm 1$, the
magnitude $R$ in
 \eq{W1osc} should be compared to the wavelength of the typical most energetic photons emitted by an electron in a trajectory $u_0(t)$.
Therefore, unless abruptly accelerated trajectories, $R$ should be
compared with $\tau_z$.  For instance, if we take $R<<\tau_z$ then
$k_x R<<1$ for values of $k_x$ where $\dot u_0(t)$'s Fourier
transform is not negligible.  Hence, in the case of coincident
trajectories ($\epsilon=+1$), we find a suppression in the
decoherence factor $\wzz$ through $\sin^2(k_x R)$ in \eq{W1osc}.
In fact, this is expectable from first principles: if the electron
oscillate in concordance in both paths and the photons wavelength
is greater than the distance between both paths, then we cannot
expect to be able to obtain the {\it which-path} knowledge from
the information imprinted into the electromagnetic field.  On the
other hand, if the electron has opposite oscillations in each path
--this is achieved by using $\epsilon=-1$--, then we should expect
to have decoherence independently of any cutoff in the spectrum of
the photons emitted by the accelerated image charges.  This is in
fact the case, since $R<<\tau_z$ and $\epsilon=-1$ imply
$\cos^2(k_x R) \approx 1$ and a value for $\wzz$ four times larger
than the case $\epsilon=0$, hence there is not suppression for
$\wzz$ in this condition.

In the other limit case, $R>>\tau_z$, we expect the photons to
carry the {\it which-path} information and to produce decoherence
independently of concordant ($\epsilon=+1$) or opposite
($\epsilon=-1)$ trajectories of the image charges in the $\hat
z$-direction.  In fact, writing the factor in the integral in
\eq{W1osc} \be \left| \frac{e^{i k_x R} \mp e^{-i k_x R}}{2}
\right|^2 = \frac{1}{2} \mp \frac{1}{4} \left( e^{i 2 k_x R} +
e^{-i 2 k_x R} \right) , \ee we see that in both cases this factor
is a constant contribution plus or minus rapid $k_x$-oscillating
contributions.  Since these oscillating factors are integrated
with the modulus square of the Fourier transform of $\dot u_0(t)$,
we find that the oscillation cancels the integral for $R>>\tau_z$.
Therefore, we have that when the distance between both paths is
large enough, the decoherence factor is independent of the value
of $\epsilon$ and equals twice its value when $\epsilon=0$: \be
\left. \wzz (\epsilon=\pm1) \right|_{R>>\tau_z} = 2  \wzz
(\epsilon=0) . \label{jjj}\ee

As an example we study here the behavior of $\wzz$ for a
particular value of $\dot u_0(t)$, which corresponds to a given
shape in the conductor's surface.   We first study analytically
the behaviour of $\wzz$ for this $\dot u_0(t)$ in the case
$\epsilon=0$, and then we study numerically its behaviour for
$\epsilon=\pm1$ and arbitrary values of $R$.

We take a $\dot u_0(t)$ which has two periods of constant
acceleration in such a way that its $t$-integral equals $\xi$, the
groove's depth. Defining $v_z \equiv\xi/\tau_z$ as the typical
velocity of the system, we have \be \dot u_0 (t) = \left\{
\begin{array}{lcl}
4v_z \, \frac{t}{\tau_z} & \qquad & t\in(0,\tau_z/2) \\
4v_z (- \frac{t}{\tau_z} + 1) & \qquad & t\in(\tau_z/2,\tau_z)
\end{array}\right. ,
\label{triangulo} \ee which, through \eq{W1osc} and \eq{main},
yields \bea W_{SP} \approx 2N \wzz (\epsilon=0)  &=&
\frac{8N}{3\pi^2} \ln (2) e^2 v_z^2 = \frac{8N}{3\pi^2} \ln (2)
e^2 (v_y \tan \theta)^2 . \label{kkk} \eea where $\theta$ is the
mean angle of the sloping surfaces.  As expected, the decoherence
increases as the slopes become more abrupt.  Although this
conclusion is achieved within a non-relativistic image-method
approximation, it is expected to be valid beyond it: an extremely
abrupt slope will induce surface currents with a greater time
dependence, which will generate decoherence through its radiation.

As it can be seen from this result, \eq{kkk}, the decoherence factor is
suppressed by a $e^2 v_z^2$ factor.  (This is an expected
suppression and it was already noticed in Ref.\cite{mpv}.)
However, the complete decoherence factor is enhanced by an $N$
factor due to the $N$ grooves in the conductor's
surface.  This enhancing factor coming from the Smith-Purcell
emitted radiation may counterbalance the $e^2 v_z^2$ suppression
and hence provide visible effects of decoherence due to a
modification of the electromagnetic boundary conditions in a
typical interference experiment.

\begin{figure}[t]
\framebox{
\includegraphics[width=.6\textwidth]{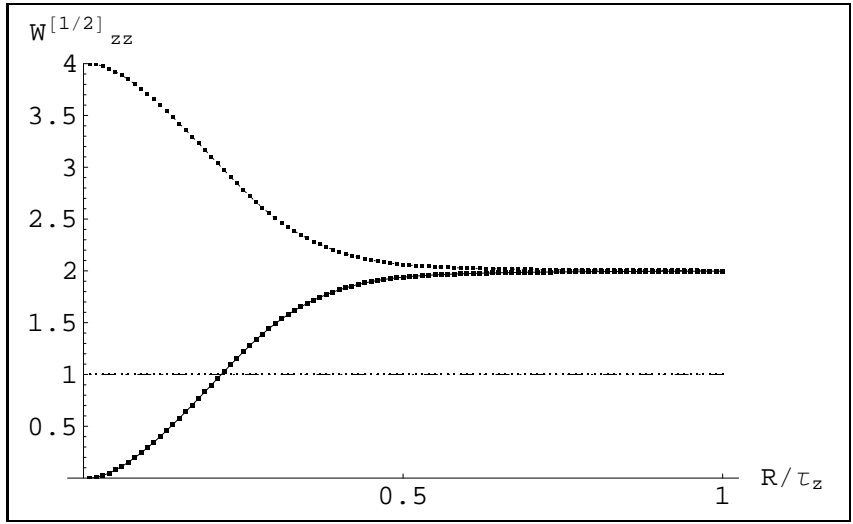} }
\caption{$\wzz$ as a function of $R/\tau_z$ for the particular
example in \eq{triangulo}.  The constant plot is for $\epsilon=0$,
whereas the other upper and lower plots correspond to
$\epsilon=-1$ and $\epsilon=+1$, respectively.  The $y$-axis is in
units of $\wzz(\epsilon=0)$, and $c=1$ is assumed.} \label{fig:triangular}
\end{figure}

The study of $\wzz$ in the $\epsilon=\pm1$ cases for arbitrary
values of $R$ is easily managed numerically, once the general
analysis
has been performed.  In Fig.~\ref{fig:triangular} we plot
the decoherence factor $\wzz$ corresponding to the velocity given
in \eq{triangulo} for $\epsilon=0,\pm1$ as a function of
$R/\tau_z$.  As expected from the above discussion,
$\wzz(\epsilon=\pm1)$ converges to $2 \wzz (\epsilon=0)$ when
$R/\tau_z>>1$ (see \eq{jjj}).

As it can be seen, the above results are independent of $z_0$ and
$d$.  This is expected in our image-method approximation, which is
valid if $z_0\ll d$. Observe that, in order to validate the
approximations, the constraint $z_0\ll R$ is also needed in the
cases $\epsilon=-1$ or $0$,  although $R$ does not appear in the
result for $\epsilon=0$ (\eq{kkk} and Fig.~\ref{fig:triangular}).

\section{Discussion and experimental perspectives}
In this article we have proposed a concrete setup to carry into
effect the idealized experiment discussed in Ref.\cite{mpv}, in
which the electron in a double slit experiment above a conducting
plane oscillates $N$ times before reaching the screen in order to
increase the decoherence factor. In the present simple setup, a
double-slit experiment for charges is performed above a grated
conducting surface. Although the electron wave packets move with
constant velocity above the grating, the image charges oscillate
many times in the vertical direction due to the presence of the
sloping surfaces. These oscillations leave an imprint in the
electromagnetic field through the Smith-Purcell radiation, and
therefore produce the sought enhancement in the decoherence
factor.

We have made several assumptions in order to simplify the
estimation of the Smith-Purcell induced overlap factor, which
turns out to be $\vert F\vert  = \exp (- C N e^2v^2)$, where $C$
is a constant of order one that depends on the details of the
conducting surface. In particular, we have assumed that the motion
of the charged particles and their images is non relativistic.
This is a  very conservative assumption, and the effect could be
larger by considering other gratings with abrupt changes in the
surface, or composed by a periodic array of conducting slabs
separated by vacuum gaps. Even within our conservative
assumptions, the effect could be large enough to be observed.  As
a concrete example, the evaluation of \eq{kkk} for the case of an
electron flying above a $N=1000$ groove's grating with
$\epsilon=0$ and a velocity and slopes such that $v_y^2 \tan^2
\theta \approx 0.1$, yields that the fringe visibility is reduced
by a factor $|F|\approx 0.42$. According to our approximations,
this may be experimentally achieved by setting $d\approx R \approx
20\mu m$, $z\lesssim 3\mu m$, and $\xi$ of the order of $1\mu m$.
This yields a total length for the conductor of about $2$ cm.

At this level of approximation ($z_0\ll d$) the decoherence factor
is independent of $z_0$ and $d$. For $z_0\gtrsim d$ we should
expect the same dependence as in the radiated power of the
non-relativistic Smith-Purcell effect, which is dominated by the
exponential factor $\exp{(-4\pi z_0/d)}$ \cite{xxx}.  However, if
we go beyond our approximation, we should also expect that the
enhancing envisaged in \eq{kkk} due to abrupt grooves becomes a
computable large factor for gratings with $\theta \sim \pi/2$.

At last, notice that an electron flying over a conductor --not
necessarily grated-- may induce an extra source of decoherence due
to dissipation of energy of the image charge in the conductor and
this could mask the Smith-Purcell decoherence.  This dissipative
effect was theoretically modelled by Anglin and Zurek
\cite{anglin} and by Machnikowski \cite{mach} and recently
measured by Sonnentag and Hasselbach \cite{sonnen}.  The
measurement of the effect still does not confirm any of the two
models.  In particular, the dependence of the effect with the
resistivity of the material ($\rho$) is quite different in the two
theoretical predictions and further experimental analysis  --as
repeating the experiment in \cite{sonnen} with different
materials-- is required in order to clarify this issue.  Using the
experimental results in \cite{sonnen} for a semiconductor with
$\rho\sim 10^{-2}\,\Omega$ m, and the $\rho$-dependence predicted
by Anglin and Zurek, the decoherence due to dissipation for the
above proposed experimental parameters using a grating of copper
at room temperature ($\rho\sim 10^{-8}\,\Omega$ m) would be
negligible. On the other hand, if we assume Machnikowski's model,
the dissipative decoherence could be comparable or greater than
$W_{SP}$. However, both models agree that if the temperature is
lowered by at least two orders of magnitude then $W_{SP}$ could be
observed, independently of possible dissipative effects.

The estimation of the overlap factor can be improved in several
directions, since we have used a crude approximation to model the
surface currents on the grating. A more rigorous approach, based
on Eq. \ref{Wcharges} with a complete evaluation of the two-point
function in the presence of the grating, could show important
deviations from our {\it proximity approximation}, in particular
for non-smooth surfaces, producing an enhancement of the effect.

\begin{acknowledgments}
We thank F.~Lombardo for carefully reading this manuscript. We
also thank D. Dalvit for comments about the validity of the
"proximity" approximation in this context. This work has been
supported by Universidad de Buenos Aires, CONICET and ANPCyT.
\end{acknowledgments}


\appendix
\section{}

In this Appendix we prove the validity of Eqs.~\ref{uno} and \ref{dos}.  To demonstrate \eq{uno} we integrate $\vec x$ and $\vec x'$ in the
expression for $W^{bb}_{zz}$, \eq{Wcdij}, and replace the velocity
$\dot u(t)$ by its expression in \eq{b}.  After performing the
change of variables $t\to t-nT_z$ in both $t$-integrals and using
$v_y \ll 1$, we obtain \bea
W^{bb}_{zz} &=& e^2 \int \frac{d^3 k}{(2\pi)^3} \frac{1}{\w} (1-\frac{k_z^2}{\w^2}) \left| \frac{e^{i k_x R} - \epsilon \, e^{-i k_x R}}{2} \right|^2 \left| \int_0^{\tau_z} dt \, \dot u_0 (t) \, e^{i\w t} \right|^2 \cdot \nonumber \\
&& \left( \sum_{n,m=0}^{2N-1} (-1)^{n+m} \, e^{i(k_y v_y + \w) T_z
(n-m)}\right) . \label{hola} \eea It is clear that there are $2N$
equal terms in the summation for which $n=m$ and, hence, give rise
to a $2N\wzz$ contribution. Now we should show that the
contribution coming from the $n\neq m$--terms is negligible in
comparison to $2N\wzz$.  Although this might look straightforward
since these other terms have an $\w$-oscillatory phase which should
suppress their contribution due to the $\w$-integration, one must
be aware that there are $\sim N^2$ of these $n\neq m$--terms.

For the sake of simplicity we present the detailed proof for the
case $\epsilon=0$, the other possible cases are analyzed below.
The relevant integral in \eq{hola} in this case is \bea \eta
\equiv e^2\sum_{n\neq m} (-1)^{n+m} \int_0^{\infty} d\tilde \w \,
\tilde \w \int_0^{\tau_z} dt\int_0^{\tau_z} dt' \dot u_0(t) \dot
u_0(t') e^{i \tilde \w (t-t'+T_z (n-m))} , \label{chau} \eea where
the change of variables $\w \to \tilde \w = k_y v_y +\w$ has been
performed and $v_y\ll1$ has been used where possible. Since the
argument of the exponential cannot be zero we can perform the
$\tilde \w$-integration in the distributional sense and get \bea
\eta &=& - e^2 \sum_{n\neq m} (-1)^{n+m} \int_0^{\tau_z} dt\int_0^{\tau_z} dt' \frac{1}{(t-t'+T_z (n-m))^2}\dot u_0(t) \dot u_0(t')\label{papel} \\
&\approx& - e^2 \sum_{n\neq m} \frac{(-1)^{n+m}}{(n-m)^2} \left(
\frac{\xi}{T_z} \right)^2 , \label{jj} \eea where in the last step
we have assumed $T_z \gg \tau_z$ in the denominator, and we have
integrated $t$ and $t'$ to obtain the height of the steep, $\xi$.
At this point we see that the $\sim N^2$ terms are combined in a
sum which can be exactly computed for large $N$ to give at leading
order, \bea \eta \approx e^2 N \frac{\pi^2}{3} \left(
\frac{\xi}{T_z} \right)^2 . \eea Having into account that in
general $\wzz\propto e^2 (\xi/\tau_z)^2$, we conclude that
$T_z\gg\tau_z$ implies $\eta \ll 2N \wzz$ and, therefore, we can
approximate \bea W^{bb}_{zz} \approx 2N \wzz, \eea as we wanted to
show.

The case $\epsilon=\pm 1$ is now easily worked out.  In this case
we need to analyze a similar $\eta$ as above, but now with a $2Re
\left(e^{i2k_x R} \right)$ factor within the $\tilde \w$ integral.
This extra-factor ends up giving a correction to the argument of
the exponential in \eq{chau} which now reads $i \tilde \w (t-t'+T_z
(n-m)+\alpha 2 R)$, where \bea \alpha = \frac{\sin\theta\, \cos
\phi}{1+v_y \,\sin\theta\, \sin \phi} \eea is an angular factor
associated to the $d^3k$-integration such that $|\alpha| \leq
1+v_y$. Therefore, if we assume $2 R<T_z(1-v_y)$ --to avoid the
possibility that the argument of the exponential vanishes--, then
we can proceed in a similar way as before and reach the same
conclusion.  Notice that for $R$ close to its larger limit, when
$|\alpha|$ takes values close to $1$ the approximation that goes
from \eq{papel} to \eq{jj} could be spoiled for the very first
terms where $n-m\sim 1$, but this does not modify our general
conclusion.

Let us now consider the inequality in \eq{dos}, $W^{ab}_{xz} +
W^{bb}_{xz} +  W^{ab}_{yz} + W^{bb}_{yz} \ll W^{bb}_{zz}$.  Once
$W^{bb}_{zz}$ has been analyzed, we can easily study the
qualitative behavior of $W^{ab}_{xz}$, $W^{bb}_{xz}$,
$W^{ab}_{yz}$ and $W^{bb}_{yz}$.  We analyze in detail
$W^{bb}_{xz}$, but the same arguments are valid as well for the
others.

Once we compute $W^{bb}_{xz}$ using \eq{Wcdij} and \eq{b}, we see
that its expression contains only one sum, instead of two.
Therefore, at difference of the $W^{zz}_{bb}$ case, there is no
cancellation of the phases that arise from the change of variables
$t\to t-n T_z$ within one of the integrals.  We have that the
relevant part of the integral in $W^{bb}_{xz}$ looks like
\bea
\sum_{n=0}^{2N-1} (-1)^n \int_0^\infty d\tilde \w \, \tilde \w \,
e^{-i(n T_Z - 2\alpha R)\tilde \w} \int dt \dot s(t) \cos(k_x s(t))
e^{i\tilde \w t}   \int dt' \dot u_0(t') e^{-i \tilde \w t'} .
\eea
Assuming that both Fourier transform within the $\tilde
\w$-integral are well behaved, we see from here that almost all
terms in the sum will have an oscillatory factor which will
suppress the contribution of the $\tilde \w$-integral.   Moreover,
the expected peak around $\tilde \w =0$ from the first Fourier
transform will be suppressed by the $\tilde \w$ factor in the
integral.  Therefore, we can expect the contribution of
$W^{bb}_{xz}$ to be negligible in comparison to $W^{bb}_{zz}$
which is enhanced by $N$.  The same arguments apply as well for
$W^{bb}_{xz}$, $W^{ab}_{yz}$ and $W^{bb}_{yz}$.


\end{document}